\documentclass[12pt]{article}
\usepackage{amsthm,amssymb,amsmath,eucal}
\usepackage[american,british]{babel}
\usepackage{graphicx,psfrag}

\newtheorem*{Th}{Theorem}

\newcommand{\dif}{\operatorname{d}}
\renewcommand{\t}{{\operatorname{t}}}
\newcommand{\I}{\operatorname{i}}

 \newcommand{\res}{\operatorname{Res}}

\newcommand{\diag}{\operatorname{diag}}
\newcommand{\bX}{{\boldsymbol X}}
\newcommand{\bn}{{\boldsymbol n}}
\newcommand{\bt}{{\boldsymbol t}}

\newcommand{\be}{{\boldsymbol e}}
\newcommand{\bc}[1]{{\boldsymbol{\mathcal{C}}^{(#1)}}}

\newcommand{\bg}{{\boldsymbol g}}
\newcommand{\bpsi}{{\boldsymbol \psi}}

\newcommand{\C}{{\mathbb C}}
\newcommand{\R}{{\mathbb R}}
\newcommand{\Z}{{\mathbb Z}}

\newcommand{\V}[1]{{\mathbb V}^{(#1)}(p,q) }
\newcommand{\W}[1]{{\mathbb V}^{(#1)}(p) }

\newcommand{\D}{{\partial}}

\begin{document}

\title{Generating\\ Quadrilateral and Circular Lattices
\\ in KP Theory\thanks{Partially supported by CICYT:
 proyecto PB95--0401}}

\author{Adam Doliwa$^{1,2}$,
Manuel Ma\~{n}as$^{3,4}$ and Luis Mart\'{\i}nez Alonso$^{4}$\\ $^1$Istituto
Nazionale di Fisica Nucleare, Sezione di Roma,\\ P-le. Aldo Moro
2, I-00185, Italy\\ $^2$Instytut Fizyki Teoretycznej, Uniwersytet
Warszawski\\ ul. Ho\.{z}a 69, 00-681 Warszawa, Poland\\
 $^3$Departamento de
Matem\'{a}tica Aplicada y Estad\'{\i}stica,\\ Escuela Universitaria de
Ingenieria T\'{e}cnica Areona\'{u}tica,\\ Universidad Polit\'{e}cnica de
Madrid,\\ E28040-Madrid, Spain.\\
$^4$Departamento de F\'{\i}sica
Te\'{o}rica II,\\ Universidad Complutense,\\ E28040-Madrid, Spain.\\ }

\maketitle

\begin{abstract}
The bilinear equations  of the $N$-component KP and BKP
hierarchies and a corresponding extended Miwa transformation allow
us to generate  quadrilateral and circular lattices from conjugate
and orthogonal nets, respectively.  The main geometrical objects
are expressed in terms of Baker functions.
\end{abstract}
\newpage

\section{Introduction}

Among the most interesting developments in Differential Geometry
that took place during the last century and the turn of it was the
study of orthogonal and conjugate nets
\cite{Darboux1,Eisenhart1,bianchi1}. The analysis of orthogonal
nets can be already found  in Lam\'{e}'s  monograph \cite{lame},
however it was Darboux who at  1910 wrote a definitive
comprehensive text \cite{Darboux2} on the subject. Darboux also
introduced the notion of  conjugate net \cite{Darboux1} and
studied it in detail. The analysis of the transformations
preserving the orthogonal or conjugate character of the net was
also developed, for an excellent review see \cite{Eisenhart2}; in
this manner, a number of transformations appeared, as for example:
Laplace, L\'{e}vy \cite{levy} and fundamental \cite{jonas,Eisenhart3}
transformations for conjugate nets, and Ribaucour \cite{ribaucour}
transformations for orthogonal nets. The iteration of two
Ribaucour transformation was performed by Demoulin \cite{demoulin}
finding the famous circularity property of these transformations.
Recently, the vectorial Ribaucour transformation \cite{qm-r} has
been presented and shown to be equivalent to the iteration of the
standard ones, see also \cite{g,gt}.

Interest from the Soliton Theory groups in this subject is not new;
indeed, the deep links among differential geometry and integrable
equations has been known for long time. In particular, Zakharov and
Manakov \cite{zm} were able to devise a $\bar\partial$-method for
solving the Darboux equations, and later an inverse scattering one
for the Lam\'{e} equations \cite{zakharov}.

Recently, the study of discrete geometry has gained quite an
audience. It was in \cite{Sauer} were the two-dimensional
quadrilateral lattice was first presented, but it was in
\cite{DQL1} were, for the multidimensional case, its  integrable
character  was shown and its transformations were presented
\cite{mds,dms,lm}. The circular lattices appeared for the first
time in \cite{circular2,circular1} and have been studied in detail
in a number of papers \cite{bobenko,cds,dms,ks,doliwa,lm2}.

In a series of works \cite{dmmms,mm} we have begun studying the
relation between these geometrical objects and integrable systems
from the point of view of the theory of the multicomponent
Kadomtsev-Petviashvili (KP) hierarchy. It is already known that
this hierarchy contains a handful of relevant soliton equations
like the Davey-Stewartson, the two-dimensional Toda lattice and
the $N$-wave resonant interaction system. Among the methods used
to describe the KP theory there are two with a remarkable
analytical character; i. e., the $\bar{\partial}$ dressing method
\cite{zm} and the bilinear identity method. The later is closely
linked with the infinite Grassmannian manifold \cite{sato,sw} and
the $\tau$-function fermion formalism \cite{djkm} (see also
\cite{kv,bk}). As for our work, we find the bilinear approach a
most convenient tool for revealing the transformation properties
of the KP dynamical objects, like Baker functions or rotation
coefficients, under the action of vertex operators. In this way we
are able to derive a series of new general identities in the KP
theory which constitute the basis of our analysis. As a
consequence,  we have proved \cite{dmmms} that the multicomponent
KP hierarchy can be regarded as the theory of iso-conjugate
deformations of conjugate nets, and that a similar result
\cite{mm} for the multicomponent BKP hierarchy and orthogonal nets
holds.

In this paper, with the help of some new identities of the KP
theory, we show how a standard vertex operator represents a
fundamental transformation of the conjugate net. Then, the
iteration of this transformation, which leads to a quadrilateral
lattice \cite{dsm}, turns out to become a Miwa-like transformation
of the KP theory \cite{m}. Such a type of transformations was
introduced in \cite{m} to generate the Hirota equation \cite{h}
from the KP hierarchy, and then  it was extensively used in a
series of papers \cite{djm} as a method of discretizing soliton
equations retaining their integrability. Our analysis provides
explicit expressions for the geometrical elements of the generated
quadrilateral lattice in terms of Baker functions. Similar results
for the multicomponent BKP hierarchy and the circular lattice are
also derived by the same means.

The layout of the paper is as follows. In next section we study the
fundamental transformation of the conjugate net and introduce the
Miwa transformation generating the corresponding quadrilateral
lattice, while in
\S 3 we perform a similar construction of the orthogonal net. In
both cases we explicitly give the form of rotation coefficients and
tangent vectors of the corresponding lattice in terms of the Baker
function.

\section{From conjugate nets to quadrilateral lattices}
\subsection{Bilinear formulation of conjugate nets}

The $N$-component KP hierarchy can be formulated in terms of the
following bilinear equation involving the $N\times N$ matrix Baker
function $\psi$ and its adjoint function $\psi^*$ \cite{djkm}
\begin{equation}\label{bilineal}
\int_{S^1}\psi(z,\bt)\psi^*(z,\bt')\dif z =0.
\end{equation}
Here $S^1:=\{ z\in\C:|z|=1\}$ is the unit circle in the complex
plane $\C$, and
$$
\bt=(\bt_1,\dots,\bt_N)\in\C^{N\raisebox{.4mm}{\scriptsize$\cdot\infty$}},\;\;
\bt_i:=(t_{i,1},t_{i,2},\dotsc)\in\C^\infty.
$$
It is also assumed that $\psi$ and $\psi^*$  can be  factorized as

\begin{equation}\label{baker}
\begin{aligned}
\psi(z,\bt)&:=\chi(z,\bt)\psi_0(z,\bt),\\
\psi^*(z,\bt)&:=\psi_0(z,\bt)^{-1}\chi^*(z,\bt),
\end{aligned}
\end{equation}
where the bare Baker functions  $\chi$ and $\chi^*$ admit Laurent
expansions on $S^1$, the first terms of which being
\begin{equation}\label{lau}
\begin{aligned}
 \chi(z)&= 1+\beta z^{-1}+{\cal
O}(z^{-2}),\quad
\\ \chi^*(z)&= 1-\beta z^{-1}+{\cal O}(z^{-2}).
\end{aligned}
\end{equation}
Notice that these Laurent series can be analitically extended from
$S^1$ to the region $|z|>1$. The \emph{vacuum} Baker function
$\psi_0$ is defined as
\[
\psi_0(z,\bt)=\exp\Big(\sum_{i=1}^N\xi(z,\bt_i)P_i\Big),\quad
\xi(z,\bt_i):=\sum_{n=1}^\infty z^nt_{i,n},
\]
with $P_i:=(\delta_{ik}\delta_{il})_{k,l=1,\dots,N}$ being the
projection matrix in the $i$-th direction.

The identity (\ref{bilineal}) encodes the Grassmannian formulation
of the KP hierarchy \cite{sato,sw} and it leads at once to  the KP
linear system of equations for $\psi$ and $\psi^*$. Let us, for
instance, outline this approach for the Baker function $\psi$. To
this end, we denote by $W$  the set of $N\times N$ matrix
functions $\varphi(z)$ such that:
\[
\int_{S^1} \varphi(z)\psi^*(z,\bt')\dif z=0,
\]
for all $\bt'$ in the definition domain of $\psi^*$. Under
appropriate conditions  $W$ belongs to an infinite-dimensional
Grassmannian manifold \cite{sato,sw}. From (\ref{bilineal}) it
follows that $W$ is a left $M_N(\C)$-module, with $M_N(\C)$ being
the ring of $N\times N$ complex matrices. Moreover, as a
consequence of (\ref{bilineal}) and (\ref{lau}) one has that for
any $\bt$:
\begin{equation}\label{sd}
W=\bigoplus_{n\geq 0}M_N(\C)
\cdot v_n(\bt),\quad v_n(z,\bt)=
\Big(\sum_{k=1}^N\frac{\partial}{\partial u_k}\Big)^n\psi(z,\bt),
\end{equation}
where $u_k:=t_{k,1}, k=1,\ldots,N$; notice also that on $S^1$ we
have
\begin{equation}\label{vn}
v_n(z)= (z^n+{\cal O}(z^{n-1}))\psi_0(z).
\end{equation}

 The linear system for $\psi$ results from the decompositions of
the time derivatives of $\psi$ in terms of $v_n$,
$n=0,\dots,\infty$. In particular, from $P_i\D\psi/\D u_k$, $i\neq
k$, one gets
\begin{equation}
\frac{\D \psi_{ij}}{\D u_k}=\beta_{ik}\psi_{kj}. \label{linear}
\end{equation}
In a similar way it follows that the adjoint Baker function
satisfies
\begin{equation}
\frac{\D \psi_{ij}^*}{\D u_k}=\beta_{kj}\psi_{ik}^*.
\label{linearad}
\end{equation}
The compatibility of either (\ref{linear}) or (\ref{linearad})
leads to the Darboux equations for the \emph{rotation
coefficients} $\beta_{ij},\; i,j=1,\ldots,N$:
\begin{equation}\label{darboux}
\frac{\D\beta_{ij}}{\D u_k}=\beta_{ik}\beta_{kj},\quad  k\neq i,j.
\end{equation}

One may recognize here the tangent vectors $\bX_i$ of the conjugate
net which can be expressed as
\begin{equation}\label{X-psi}
\bX_i=\int_{S^1}\bpsi_i(z)f(z)\dif z
\end{equation}
where $\bpsi_i$ denotes the $i$-th row of $\psi$ and $f=\diag
(f_1,\dots,f_N)$ is a suitable diagonal distribution matrix. In a
similar way, the Lam\'{e} coefficients can be constructed as follows
\begin{equation} \label{H-psi}
H_i(\bt)=\sum_{j=1}^N\int_{S^1}h_j(z)\psi^*_{ji}(z,\bt)\dif z,
\quad i=1,\cdots, N,
\end{equation}
for suitable distributions $\{h_j\}_{j=1,\dotsc,N}$. Obviously the
higher times $\bt_2,\bt_3,\dots$ can be considered as
iso-conjugate deformation parameters \cite{dmmms}.

An essential element in the subsequent discussion  is  the  vertex
operator
\[
\V{i}f(\bt):=f(\bt+([p]-[q])\be_i)
\]
where $\{\be_i\}_{i=1}^N$ is the canonical basis of $\C^N$ and
\[
[z]:=\bigg(\frac{1}{z},\frac{1}{2z^2},\frac{1}{3z^3},\dots\bigg).
\]
As it is proved below, from a geometrical point of view this
operator represents a fundamental transformation of the conjugate
net.

\subsection{Action of vertex operators on Baker functions}

The bilinear identity (\ref{bilineal}) is a useful tool for
obtaining the transformation properties of the rotation
coefficients and the Baker functions under the action of vertex
operators. We first observe that given $p,q\in \C$ such that
$|p|,|q|>1$, then for all $z\in S^1$ it follows that
\begin{equation}\label{vpsi}
\V{i}\psi(z,\bt)=[\V{i}
\chi(z,\bt)]\psi_0(z,\bt)\Big[1-\Big(1-\frac{p}{q}\Big)\frac{z}{z-p}P_i\Big],
\end{equation}
so that $\V{i}\psi(z)$ can be meromorphically extended to the
region with $|z|\geq 1$ being the unique singularity  a simple
pole located at $z=p$. The corresponding residue satisfies
\begin{equation}\label{ress}
\res_p[\V{i}\psi](1 -P_i)=0.
\end{equation}
In this way, if  we perform the substitutions
$\bt\to\bt+([p]-[q])\be_i$ and $\bt'\to\bt$  in (\ref{bilineal})
and evaluate the resulting integral, the only non-vanishing
contributions come from residues at $z=\infty$ and $z=p$. Hence we
find
\begin{multline}\label{residues}
 -p\Big(1-\frac{p}{q}\Big)P_i+ [\V{i}\beta]\Big[1-\Big(1-\frac{p}{q}\Big)P_i\Big]-
  \Big[1-\Big(1-\frac{p}{q}\Big)P_i\Big]\beta\\=\res_p[\V{i}\psi]\psi^*(p).
\end{multline}

In the following diagram we show the polar structure and the
contour of integration in the Riemann sphere
 \psfrag{p}{$p$}  \psfrag{s}{$S^1$}
\psfrag{I}{$\infty$}\psfrag{c}{$\boldsymbol{\bar{\mathbb C}}$}
\begin{center}
\includegraphics[width=7cm]{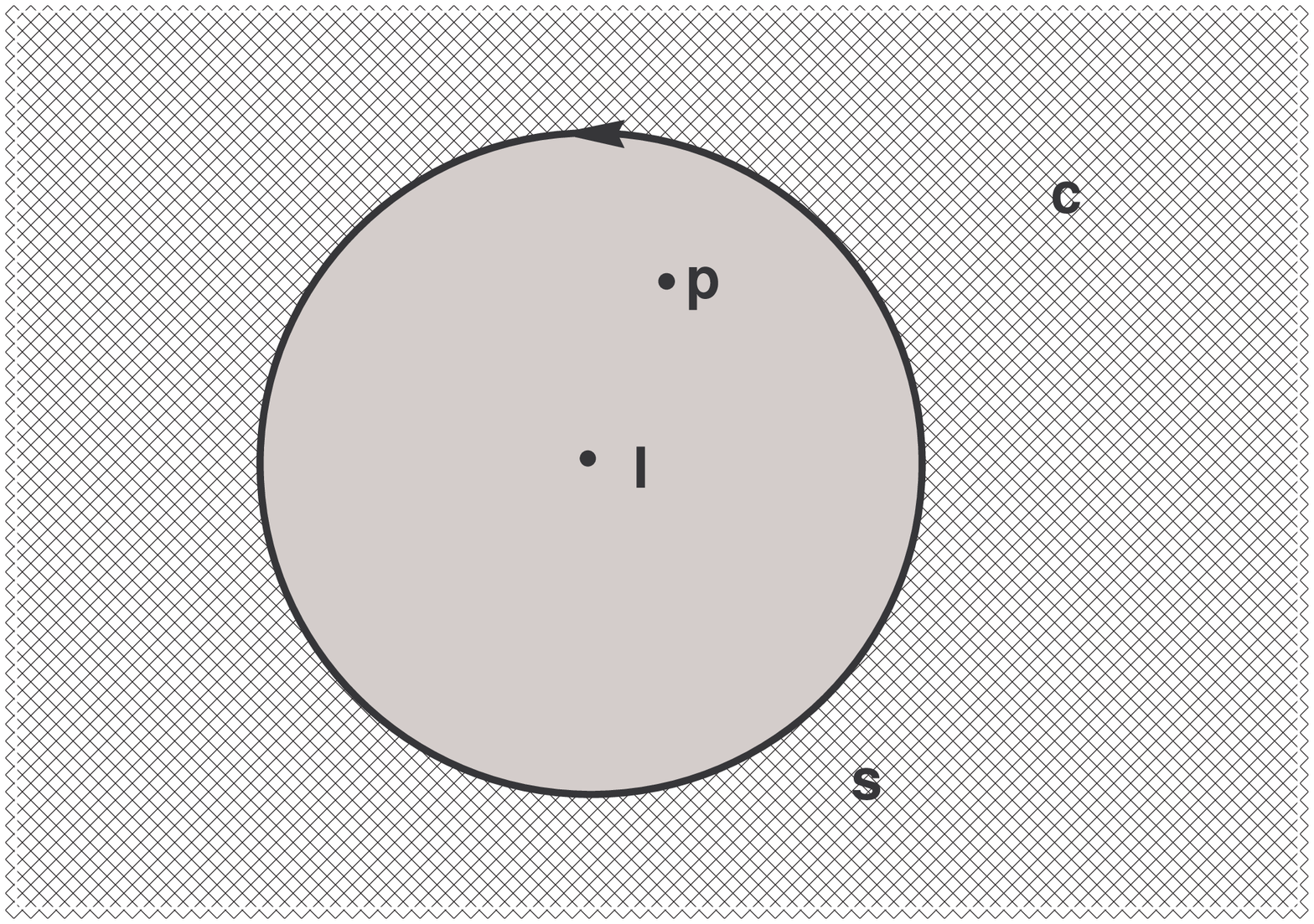}
\end{center}

The components of this matrix equation can be obtained by
multiplying (\ref{residues}) to the left and to the right by the
projectors $P_j$, $i=1,\dots,N$. Thus, one finds the following
identities for the action of $\V{i}$ on the rotation coefficients
\begin{align}
  \V{i}\beta_{ii} & =\beta_{ii}+q\Big(1-\frac{p}{q}\Big)+\frac{q}{p}
  \res_p(\V{i}\psi_{ii})\psi_{ii}^*(p),\label{beta'ii}\\
  \frac{q}{p}\V{i}\beta_{ij} &
  =\beta_{ij}+\frac{q}{p}\res_p(\V{i}\psi_{ii})\psi_{ij}^*(p),\quad i\neq j,\label{beta'ij}\\
\frac{p}{q}\V{i}\beta_{ji} &
  =\beta_{ji}+\res_p(\V{i}\psi_{ji})\psi_{ii}^*(p),\quad i\neq j,\label{beta'ji}\\
\V{i}\beta_{jk} &
  =\beta_{jk}+\res_p(\V{i}\psi_{ji})\psi_{ik}^*(p),\quad \text{$i,j$ and $k$ different.}
  \label{beta'jk}
\end{align}

Our next aim is to derive similar identities for the Baker function
$\psi$. To this end we consider the function
\begin{equation}\label{f}
F:=P_j\Big[\dfrac{\res_p(\V{i}\psi)}{\res_p(\V{i}\psi_{ii})}
  -1\Big]\V{i}\psi,
\end{equation}
and observe that its unique possible singularity in the region
$|z|>1$ might be located at $z=p$. However, it is easy to see that
(\ref{ress}) implies that $z=p$ is a removable singularity of $F$.
Therefore, the function $F$ can be analytically extended from
$S^1$ to the region $|z|>1$. Furthermore, from (\ref{vpsi}) it
follows that
\[
\V{i}\psi(z)\sim
(1-\Big(1-\frac{p}{q}\Big)P_i+\mathcal{O}(z^{-1}))\psi_0(z),\quad
z\to\infty,
\]
so that the first terms of the Laurent expansion of $F$ on $S^1$
are
$$ F= P_j\Big[\displaystyle\frac{\res_p(\V{i}\psi)}{
 \res_p(\V{i}\psi_{ii})}
-1\Big](1-\Big(1-\frac{p}{q}\Big)P_i+\mathcal{O}(z^{-1}))\psi_0(z).
$$ Hence from the uniqueness of the Baker function in the
Grassmannian element $W$ we deduce
 \[
 P_j\Big[\frac{\res_p(\V{i}\psi)}{\res_p(\V{i}\psi_{ii})}-1\Big]\V{i}\psi(z) 
 = P_j\Big[\frac{p}{q}\frac{\res_p(\V{i}\psi)}{\res_p(\V{i}\psi_{ii})}
-1\Big]\psi(z).
\]
In terms of the rows $\bpsi_j$, this identity can be rewritten as
\begin{equation}\label{bpsi'}
  \V{i}\bpsi_j=
  \bpsi_j+\frac{\res_p(\V{i}\psi_{ji})}{\res_p(\V{i}\psi_{ii})}
  \Big(\V{i}\bpsi_i-\frac{p}{q}\bpsi_i\Big),\quad j\neq i.
\end{equation}
By noticing now that (\ref{vpsi}) implies that $\V{i}\psi(q)P_i=0$,
we get from (\ref{bpsi'}) the following useful identity
\begin{equation}\label{psiji}
  \frac{\psi_{ji}(q)}{\res_p(\V{i}\psi_{ji})}=
  \frac{p}{q}\frac{\psi_{ii}(q)}{\res_p(\V{i}\psi_{ii})},\quad j\neq i.
\end{equation}

\subsection{From vertex operators to fundamental transformations}

With the help of the above identities it is now straightforward to
formulate a representation of a fundamental transformation  as a
vertex operator.  We first prove that the following operator acting
on the rotation coefficients
\begin{align*}
  \mathcal{F}^{(i)}(\beta_{ij}) & :=\frac{q}{p}\V{i}\beta_{ij}, \quad i\neq j\\
  \mathcal{F}^{(i)}(\beta_{ji}) & :=\frac{p}{q}\V{i}\beta_{ji},\quad i\neq j\\
\mathcal{F}^{(i)}(\beta_{jk}) & :=\V{i}\beta_{jk},\quad
\text{$i,j$ and $k$ different }.
\end{align*}
is a fundamental transformation.

In view of (\ref{beta'ii}-\ref{beta'jk}) and taking (\ref{psiji})
into account, the action of $\mathcal{F}^{(i)}$ can be expressed as
\[
\mathcal{F}^{(i)}(\beta_{jk})=\beta_{jk}-\frac{v^{(i)}_j{v^{(i)}_k}^*}{\Omega^{(i)}},
\]
where
\begin{equation}\label{data}
\begin{aligned}
  v^{(i)}_j & :=\psi_{ji}(q), \\
  {v^{(i)}_k}^* & :=\psi_{ik}^*(p),\\
  \Omega^{(i)}&:=-\frac{p}{q}\frac{\psi_{ii}(q)}{\res_p(\V{i}\psi_{ii})}.
\end{aligned}
\end{equation}
Obviously, $v^{(i)}_j$ and ${v^{(i)}_k}^*$ are solutions of the
linear systems (\ref{linear}) and (\ref{linearad}). Therefore, in
order  to conclude that $\mathcal{F}^{(i)}$ is a fundamental
transformation with data given by (\ref{data}) we have to show
that $\Omega^{(i)}$ is a \emph{potential}, i. e.:
\[
\frac{\D\Omega^{(i)}}{\D u_j}=v^{(i)}_j{v^{(i)}_j}^*.
\]
 For example, for $j\neq i$, the use of (\ref{linear}),
(\ref{psiji}) and (\ref{beta'ij}) leads to the desired result. The
$i$-th derivative must be treated with more care, in fact, we have
to start with the alternative expressions
\[
\Omega^{(i)}=\frac{\psi_{ji}(q)}{\res_p(\V{i}\psi_{ji})},\quad
j\neq i;
\]
take the $i$-th derivative and use (\ref{beta'ji}).

As for the fundamental transformation acting on the tangent vectors
of the net, it is given by
\begin{align*}
  \mathcal{F}^{(i)}(\bX_i) & :=\frac{q}{p}\V{i}\bX_i, \\
  \mathcal{F}^{(i)}(\bX_j) & :=\V{i}\bX_j,\quad j\neq i,
\end{align*}
which can be written as
\[
\mathcal{F}^{(i)}(\bX_j)=\bX_j-\frac{v^{(i)}_j}{\Omega^{(i)}}\bc{i},
\]
where
\[
\bc{i}:=\frac {\V{i}
\bX_i-\dfrac{p}{q}\bX_i}{\res_p(\V{i}\psi_{ii})},
\]
is clearly the  Combescure vector of this fundamental geometrical
transformation: in  fact  a simple check shows that
\[
\frac{\D\bc{i}}{\D u_j}=\bX_j {v^{(i)}_j}^*.
\]
For $j\neq i$ one just take the derivative and then uses
\eqref{X-psi}, (\ref{linear}), (\ref{psiji}) and (\ref{beta'ij}).
The $j=i$ case requires the use of the following expression for
$\bc{i}$ ( it derives from (\ref{psiji}))
\[
\bc{i}=\displaystyle\frac{\V{i}\bX_j-\bX_j}{\res_p(\V{i}\psi_{ji})}
\]
and formulae \eqref{X-psi}, (\ref{linear})  and (\ref{beta'ji}).

\subsection{Miwa transformations
 and the quadrilateral lattice}

We are going to see how the quadrilateral lattice is generated from
the conjugate net by discretizing variables in the Baker functions
by means of a Miwa like transformation \cite{lm}
\begin{equation}\label{disc}
\begin{aligned}
  \Psi(z,\bt,\bn) & :=\psi(z,\bt+\bn([p]-[q])), \\
  \Psi^*(z,\bt,\bn) & :=\psi^*(z,\bt+\bn([p]-[q])),
\end{aligned}
\end{equation}
where $\bn=n_1\be_1+\dots+n_N\be_N\in\Z^N$. Notice that from the
point of view of our above discussion  the  discrete variables in
(\ref{disc}) are introduced by a composition of fundamental
transformations acting on tangent vectors of the conjugate net.
Therefore, it is known \cite{dsm}  that as a consequence the
corresponding lattice is a quadrilateral lattice with suitable
re-normalized tangent vectors given by the Combescure vectors
$\{\bc{i}\}_{i=1}^N$. However, finding the \emph{quadrilateral}
rotation coefficients $Q_{ij}$ requires to consider the iteration
$\mathcal{F}^{(i)}\circ\mathcal{F}^{(j)}=:\mathcal{F}^{(i,j)}$.
For that aim we consider the composed vertex operator
$\V{i,j}=\V{i}\circ\V{j}$.

Since
\begin{equation}\label{vpsij}
\V{i,j}\psi(z,\bt)=[\V{i,j}
\chi(z,\bt)]\psi_0(z,\bt)\Big[1-\Big(1-\frac{p}{q}\Big)\frac{z}{z-p}(P_i+P_j)\Big],
\end{equation}
we have a simple pole appearing at $z=p$. In fact this equation is
(\ref{vpsi}) after the replacement of the projector $P_i$ by the
projector $P_i+P_j$. Thus by substituting
$\bt\to\bt+([p]-[q])(\be_i+\be_j)$ and $\bt'\to\bt$ in
(\ref{bilineal}), the only non-vanishing contributions to the
integral come from $z=\infty$ and $z=p$,so
\begin{multline}\label{residuesij}
 -p\Big(1-\frac{p}{q}\Big)(P_i+P_j)+ [\V{i,j}\beta]\Big[1-\Big(1-\frac{p}{q}\Big)(P_i+P_j)\Big]-
  \Big[1-\Big(1-\frac{p}{q}\Big)(P_i+P_j)\Big]\beta\\=\res_p[\V{i,j}\psi]\psi^*(p).
\end{multline}
We introduce now some convenient notation
\begin{gather*}
  B :=\begin{pmatrix}
    \beta_{ii} & \beta_{ij} \\
    \beta_{ji} & \beta_{jj} \
  \end{pmatrix},\quad  b_k:=\begin{pmatrix}
    \beta_{ik} \\
    \beta_{jk} \
  \end{pmatrix},\quad  \tilde b_k:=\begin{pmatrix}
    \beta_{ki} & \beta_{kj} \
  \end{pmatrix}, \\
  \Phi := \begin{pmatrix}
    \psi_{ii} & \psi_{ij} \\
    \psi_{ji} & \psi_{jj} \
  \end{pmatrix},\quad  \phi_k:=\begin{pmatrix}
    \psi_{ik} \\
    \psi_{jk} \
  \end{pmatrix},\quad \tilde\phi_k:=\begin{pmatrix}
    \psi_{ki} & \psi_{kj} \
  \end{pmatrix},\
\end{gather*}
which allows us to express the components of the matrix equation
(\ref{residuesij}) as
\begin{align}
  \V{i,j}B & =B+q\Big(1-\frac{p}{q}\Big)+\frac{q}{p}
  \res_p(\V{i,j}\Phi)\Phi^*(p),\label{B'}\\
  \frac{q}{p}\V{i,j}b_k &
  =b_k+\frac{q}{p}\res_p(\V{i,j}\Phi)\phi_k^*(p),\quad \text{$i,j$ and $k$ different},\label{b'k}\\
\frac{p}{q}\V{i,j}\tilde b_k &
  =\tilde b_k+\res_p(\V{i,j}\tilde\phi_k)\Phi^*(p),\quad \text{$i,j$ and $k$ different},\label{tildeb'k}\\
\V{i,j}\beta_{kl} &
  =\beta_{kl}+\res_p(\V{i,j}\tilde\phi_k)\phi_k^*(p),\quad \text{$i,j,k$ and $l$ different.}
  \label{beta'kl}
\end{align}

We aim now to get similar identities for the Baker function. From
(\ref{vpsij}) it is clear that
\begin{equation}\label{res0}
\res_p(\V{i,j}\psi)= \res_p(\V{i,j}\psi)(P_i+P_j),
\end{equation}
that implies
\begin{equation}\label{res2} \res_p(
  P_k[\res_p(\V{i,j}\psi)\rho_{ij}[(\res_p(\V{i,j}\Phi))^{-1}]-1]\V{i,j}\psi)=0,
\end{equation}
where $\rho_{ij}:M_{2\times2}\to M_{N\times N}$ with
$\rho_{ij}[(m_{ab})_{a,b=i,j}]=(m_{ab}\delta_{ak}\delta_{bl})_{\substack{k,l=1,\dots,N\\a,b=i,j}}$
is the canonical embedding of $2\times 2$ matrices into $N\times
N$ matrices in the cross of the $i$-th and $j$-th columns and
rows.

The same arguments as in the previous subsection lead us to the
identity
\begin{multline*}
P_k[\res_p(\V{i,j}\psi)\rho_{ij}[(\res_p(\V{i,j}\Phi))^{-1}]-1]\V{i,j}\psi\\
 = P_k\Big[\frac{p}{q}\res_p(\V{i,j}\psi)\rho_{ij}[(\res_p(\V{i,j}\Phi)^{-1}]-
 1\Big]\psi,
\end{multline*}
which in terms of the rows of $\psi$ reads
\begin{multline}\label{bpsi'ij}
  \V{i,j}\bpsi_k=
  \bpsi_k+\res_p(\V{i,j}\tilde\phi_{k})\\ \quad \times[\res_p(\V{i,j}\Phi)]^{-1}
  \begin{pmatrix}
    \V{i,j}\bpsi_i-\dfrac{p}{q}\bpsi_i \\[.5cm]
    \V{i,j}\bpsi_j-\dfrac{p}{q}\bpsi_j \
  \end{pmatrix}.
\end{multline}
Finally, from (\ref{vpsij}) it follows that
$\V{i,j}\psi(q)(P_i+P_j)=0$, that together with (\ref{bpsi'ij})
gives
\begin{equation*}
  \tilde\phi_k(q)\Phi(q)^{-1}=
  \frac{p}{q}\res_p(\V{i,j}\tilde\phi_k)[\res_p(\V{i,j}\Phi)]^{-1},\quad k\neq i,j.
\end{equation*}

\paragraph{Identification with the iterated fundamental transformation}

As it is known \cite{dsm}  the vectorial fundamental
transformation
 is just a composition of two different
fundamental transformations. It turns out that the following
operator acting on  the rotation coefficients
\begin{align*}
  \mathcal{F}^{(i,j)}(b_k) & :=\frac{q}{p}\V{i,j}b_k, \quad k\neq i,j,\\
  \mathcal{F}^{(i,j)}(\tilde b_k) & :=\frac{p}{q}\V{i,j}
  \tilde b_k,\quad k\neq i,j,\\
\mathcal{F}^{(i,j)}(\beta_{kl}) & :=\V{i,j}\beta_{k,l},\quad
\text{$i,j,k$ and $l$ different, }
\end{align*}
is a iterated (or  vectorial) fundamental transformation. Indeed,
from (\ref{B'}-\ref{beta'kl}) we can write
\[
\mathcal{F}^{(i,j)}(\beta_{kl})=\beta_{kl}-\tilde\phi_k(q){\Omega^{(i,j)}}^{-1}
\phi^*_l(p),
\]
where
\[
\Omega^{(i,j)}:=-\frac{p}{q}[\res_p(\V{i,j}\Phi)]^{-1}\Phi(q).
\]
As in the previous section is not difficult to check that
$\Omega^{(i,j)}$ is a potential matrix; i. e.,
\[
\frac{\D\Omega^{(i,j)}}{\D u_k}=\phi_k^*(p)\tilde\phi_k(q).
\]

As for the vectorial fundamental transformation of the tangent
vectors, it takes the form
\begin{align*}
  \mathcal{F}^{(i,j)}(\bX_k) & :=\frac{q}{p}\V{i,j}\bX_k,\quad k=i,j, \\
  \mathcal{F}^{(i,j)}(\bX_k) & :=\V{i,j}\bX_k,\quad k\neq i,j,
\end{align*}
Observe that according to (\ref{bpsi'ij}) we have
\[
\mathcal{F}^{(i,j)}(\bX_k)=\bX_k-\tilde\phi_k{\Omega^{(i,j)}}^{-1}
[\res_p(\V{i,j}\Phi)]^{-1}\begin{pmatrix}
  \V{i,j}\bX_i-\dfrac{p}{q}\bX_i\\[.5cm]
  \V{i,j}\bX_j-\dfrac{p}{q}\bX_j
\end{pmatrix}.
\]

Now, we observe that
\[
\rho_{ij}[(\res_p(\V{i,j}\Phi))^{-1}]\V{i,j}\psi- \sum_{k=i,j}
P_k\frac{\V{k}\psi}{\res_p(\V{k}\psi_{kk})}
\]
is an analytical function in the region $|z|\geq 1$ up to a
possible simple pole at $z=p$; however, we see that its residue
there vanishes and therefore $z=p$ is a removable singularity.
Thus, the above function is holomorphic outside the unit circle.
Taking into account its asymptotic expansion at $z=\infty$ and the
uniqueness of the Baker function in the corresponding element in
the Grassmannian we deduce:
\begin{multline*}
 \rho_{ij}[(\res_p(\V{i,j}\Phi))^{-1}]\V{i,j}\psi- \sum_{k=i,j}
P_k\frac{\V{k}\psi}{\res_p(\V{k}\psi_{kk})}
 \\=
\frac{p}{q}\Big(\rho_{ij}[(\res_p(\V{i,j}\Phi))^{-1}]-
\sum_{k=i,j}\frac{ P_k}{\res_p(\V{k}\psi_{kk})}\Big)\psi.
\end{multline*}
From the rows of this matrix relation we obtain
\[
[\res_p(\V{i,j}\Phi)]^{-1}\begin{pmatrix}
  \V{i,j}\bX_i-\dfrac{p}{q}\bX_i\\[.5cm]
  \V{i,j}\bX_j-\dfrac{p}{q}\bX_j
\end{pmatrix}=
  \begin{pmatrix}
  \bc{i} \\
  \bc{j}
\end{pmatrix},
\]
and we have
\[
\mathcal{F}^{(i,j)}(\bX_k)=\bX_k-\tilde\phi_k{\Omega^{(i,j)}}^{-1}
\begin{pmatrix}
  \bc{i} \\
  \bc{j}
\end{pmatrix}.
\]
The above formula is just the vectorial fundamental transformation
for the tangent vectors of the conjugate net.

Now, since the fundamental transformation of the Combescure
vectors satisfies
\[
\mathcal{F}^{(j)}(\bc{i})=\bc{i}-\frac{\Omega^{(i,j)}_{ij}}{\Omega^{(i,j)}_{jj}}
\bc{j}, \quad i\neq j,
\]
then by performing the discretization (\ref{disc}) the
transformation $\mathcal{F}^{(j)}$ becomes the shift $T_j$ in the
$n_j$ discrete variable. In this way we obtain the desired
quadrilateral lattice equations
\[
\Delta_j\bc{i}=(T_jQ_{ij})\bc{j},\quad i\neq j,
\]
with
\[
T_jQ_{ij}:=\frac{[\res_p(\V{i,j}\psi_{jj})]\psi_{ij}(q)-
[\res_p(\V{i,j}\psi_{ij})]\psi_{jj}(q)}{[\res_p(\V{i,j}\psi_{ji})]\psi_{ij}(q)-
[\res_p(\V{i,j}\psi_{ii})]\psi_{jj}(q)},\quad i\neq j,
\]
being the shifted rotation coefficients of the lattice.
\newpage

We can summarize the results of this section in

\begin{Th}
The Miwa transformation generates a quadrilateral lattice with
discrete rotation coefficients and  renormalized tangent vectors
given by
\begin{align*}
T_jQ_{ij}&= \frac{[\res_p(\V{i,j}\psi_{jj})]\psi_{ij}(q)-
[\res_p(\V{i,j}\psi_{ij})]\psi_{jj}(q)}{[\res_p(\V{i,j}\psi_{ji})]\psi_{ij}(q)-
[\res_p(\V{i,j}\psi_{ii})]\psi_{jj}(q)},\quad i\neq j, \\
\bc{j}&=
\frac{1}{\res_p(\V{j}\psi_{jj})}\Big(\V{j}\bX_j-\dfrac{p}{q}\bX_j\Big).
\end{align*}
\end{Th}

\section{From orthogonal nets to circular lattices}

\subsection{Bilinear formulation of orthogonal nets}

The $N$-component BKP hierarchy can be formulated in terms of the
following bilinear equation \cite{BKP,mm}
\begin{equation}\label{bilineal-b}
\int_{S^1}\psi(z,\bt)\psi^\t(-z,\bt')\frac{\dif z}{2\pi \I
z} =G(\bt)G(\bt'),\quad G(\bt)^\t=G(\bt)^{-1},
\end{equation}
where
$$
\bt:=(\bt_1,\dots,\bt_N),\; \bt_i=(t_{i,1},t_{i,3},\dots).
$$
It is assumed that the Baker function $\psi$  admits a
factorization
\begin{equation}\label{baker-b}
\psi(z,\bt):=\chi(z,\bt)\psi_0(z,\bt),
\end{equation}
and that $\chi$ has a Laurent expansion on $S^1$ of the form
\begin{equation}
\label{asym-b}
 \chi(z)= 1+\beta z^{-1}+{\cal O}(z^{-2})
\end{equation}
and
\[
\psi_0(z,\bt)=\exp(\sum_{i=1}^N\xi(z,\bt_i))P_i),\quad
\xi(z,\bt_i)=\sum_{i=1}^\infty z^{2n-1}t_{i,2n-1},
\]
As a consequence of (\ref{bilineal-b}) one finds that the rotation
coefficients $\beta_{ij}$ and the Baker function satisfy
(\ref{linear}) and (\ref{darboux}). Moreover one can identify
$G(\bt)$ and $\psi(z,\bt)|_{z=0}$ \cite{mm}, thus
$\{\bg_i\}_{i=1}^N$ is an orthonormal frame and therefore the
conjugate net is orthogonal; i. e., we have the Lam\'{e} equations for
the rotation coefficients
\begin{gather}
\frac{\partial\beta_{ij}}{\D u_k}-\beta_{ik}\beta_{kj}=0,\;\;
i,j,k=1,\dotsc, N,\; \text{with $i,j,k$ different},\label{lame1}
\\
\frac{\partial\beta_{ij}}{\D u_i}+ \frac{\partial\beta_{ji}}{\D
u_j}+ \sum_{\substack{k=1,\dotsc,N\\ k\neq i,j}}
\beta_{ki}\beta_{kj}=0,\quad i,j=1,\dotsc,N,\;i\neq
j.\label{lame2}
\end{gather}

There is again a vertex operator which plays a fundamental role in
the analysis of geometrical transformations for the orthogonal net.
It is given by
\[
\W{i}f(\bt):=f(\bt+[p]\be_i)
\]
where now
\[
[z]:=2\bigg(\frac{1}{z},\frac{1}{3z^3},\dots\bigg).
\]
Indeed, as it is proved in \cite{lame}, it follows that $\W{i}$ is
related to the well-known Ribaucour transformation in the form
\[
{\cal R}^{(i)}(\bg_j):=(-1)^{\delta_{ij}}\W{i}\bg_j= \bg_j-
\frac{2 \psi_{ji}(-p)}{|\bc{i}|^2}\bc{i},
\]
where the Combescure vector is given by
\[
\bc{i}:=\sum_{k=1}^N\psi_{ki}(-p)\bg_k.
\]
\subsection{Miwa transformations and the circular lattice}

The circular lattice can be obtained from the the orthogonal net by
discretizing variables in the Baker function through the Miwa
transformation
\begin{equation}\label{disc1}
  \Psi(z,\bt,\bn):=\psi(z,\bt+\bn[p]).
\end{equation}
This property derives from the fact that (\ref{disc1}) is a
composition of Ribaucour transformations acting on tangent vectors
of the orthogonal net, and therefore, as it was shown by Demoulin
\cite{demoulin}, it generates a circular lattice. The suitably
renormalized tangent vectors are just the Combescure vectors
$\{\bc{i}\}_{i=1}^N$. However,  the computation of the
 coefficients $Q_{ij}$ of the corresponding circular lattice requires the
use of the iteration
$:\mathcal{R}^{(i,j)}=\mathcal{R}^{(i)}\circ\mathcal{R}^{(j)}$.
Thus we are lead to  consider the composed vertex operator
$\W{i,j}=\W{i}\circ\W{j}$ and to perform the substitutions
$\bt\to\bt+[p](\be_i+\be_j)$ and $\bt'\to\bt$ in
(\ref{bilineal-b}).

 Firstly, notice that
\begin{equation}\label{wpsij}
\W{i,j}\psi(z)=[\W{i,j}
\chi(z)]\psi_0(z)\Big[1-\frac{2z}{z-p}(P_i+P_j)\Big].
\end{equation}
This implies
$$
\res_p(\W{i,j}\psi)=\res_p(\W{i,j}\psi(z))(P_i+P_j).
$$
Moreover, the only non-vanishing contributions in
(\ref{bilineal-b})
  come from $z=\infty$ and $z=p$, as a consequence we have
 \begin{equation}\label{residuo-b}
 1-2(P_i+P_j)-\res_p(\W{i,j}\psi)
 \psi^\t(-p)=[\W{i,j}G] G^\t.
 \end{equation}
 The orthogonal character of the right-hand side in the
 above formula gives
\begin{multline*}
(1-2(P_i+P_j))\psi(-p))\res_p(\W{i,j}\psi^\t)+
\res_p(\W{i,j}\psi) \psi^\t(-p)(1-2(P_i+P_j))\\=
 \res_p(\W{i,j}\psi)\psi^\t\psi
\res_p(\W{i,j}\psi^\t).
\end{multline*}

We use the same notation for the fundamental transformation as in
the previous section , now:
\[
\Omega^{(i,j)}:=-[\res_p(\W{i,j}\Phi)]^{-1}\Phi(-p).
\]
 This equation has two implications:
\begin{gather*}
\Omega^{(i,j)}+{\Omega^{(i,j)}}^\t=\sum_{k=1}^N\tilde\phi_k^\t(-p)\tilde\phi_k(-p),\\
\tilde\phi_k(-p)=\res_p(\W{i,j}\tilde\phi_k)\Omega^{(i,j)}.
\end{gather*}
On the other hand, from (\ref{residuo-b}) it follows
\begin{align*}
  \mathcal{R}^{(i,j)}(\bg_k)&:=
  (-1)^{\delta_{ik}+\delta_{jk}}\W{i,j}\bg_k=\bg_k-(-1)^{\delta_{ik}+\delta_{jk}}
\res_p(\W{i,j}\tilde\phi_k)\begin{pmatrix}
  \bc{i} \\
  \bc{j}
\end{pmatrix} \\
  & =
\bg_k-\tilde\phi_k(-p){\Omega^{(i,j)}}^{-1}\begin{pmatrix}
  \bc{i} \\
  \bc{j}
\end{pmatrix},
\end{align*}
as it should be. Observe that we have
\begin{align*}
  \Omega_{ii}^{(i,j)} & =\dfrac{1}{2}\sum_{k=1}^N \psi_{k,i}^2(-p)=\dfrac{1}{2}|\bc{i}|^2,\\
 \Omega_{jj}^{(i,j)} & =\dfrac{1}{2}\sum_{k=1}^N \psi_{k,j}^2(-p)=\dfrac{1}{2}|\bc{j}|^2,\\
 \Omega_{ij}^{(i,j)}+\Omega_{ji}^{(i,j)}&=\sum_{k=1}^N
 \psi_{k,i}(-p)\psi_{k,j}(-p)=\bc{i}\cdot\bc{j}.
\end{align*}
Now, the Ribaucour transformation of the Combescure vector is
\[
\mathcal{R}^{(j)}(\bc{i})=\bc{i}-2\frac{\Omega^{(i,j)}_{ij}}{|\bc{j}|^2}\bc{j}.
\]
Hence, provided the discretization (\ref{disc1}) is applied the
transformation  $\mathcal{R}^{(j)}=T_j$ becomes the shift in the
$n_j$ discrete variable and we get the quadrilateral lattice
equations
\[
\Delta_j\bc{i}=(T_jQ_{ij})\bc{j},
\]
where the rotation coefficients are given by:
\[
T_jQ_{ij}:=2\frac{\res_p(\W{i,j}\psi_{jj})\psi_{ij}(-p)-
\res_p(\W{i,j}\psi_{ij})\psi_{jj}(-p)}{\sum_{k=1}^N\psi_{kj}^2(-p)}.
\]
Moreover, the circularity property follows from the identity
\[
  \bc{i}\cdot\mathcal{R}^{(i)}(\bc{j})+\bc{j}\cdot
  \mathcal{R}^{(j)}(\bc{i})
  =2\bc{i}\cdot\bc{j}-2(\Omega_{12}^{(i,j)}+\Omega_{12}^{(i,j)})=0;
\]
  i.e,
\[
\bc{i}\cdot T_i(\bc{j})+\bc{j}\cdot
  T_j(\bc{i})=0.
\]

Summarizing:

\begin{Th}
The Miwa transformation generates a circular lattice with discrete
rotation coefficients and  renormalized tangent vectors given by
\begin{align*}
T_jQ_{ij}&= 2\frac{\res_p(\W{i,j}\psi_{jj})\psi_{ij}(-p)-
\res_p(\W{i,j}\psi_{ij})\psi_{jj}(-p)}{\sum_{k=1}^N\psi_{kj}^2(-p)},
\\  \bc{j}&=\sum_{k=1}^N\psi_{kj}(-p)\bg_k.
\end{align*}
\end{Th}

\end{document}